\documentclass[a4paper,11pt,twoside]{article}
 \pdfoutput=1
\usepackage[english]{babel}
\makeatletter

\usepackage{geometry,graphicx,color}
\usepackage{amsmath,amssymb}
\usepackage[pdftex]{hyperref}
\usepackage{enumerate}

\usepackage[utf8]{inputenc}
\usepackage{amsthm}
\usepackage{mathrsfs}
\usepackage{lmodern}
\usepackage{amsmath}
\usepackage{amsfonts}
\usepackage{amssymb}
\usepackage{bm}
\numberwithin{equation}{section}
\newcommand{\institute}[1]{\newcommand{\@institute}{#1}}
\renewcommand{\maketitle}{
\vspace*{0.5\baselineskip}
{
\center\LARGE\noindent\@title\par
}%
\vspace{1.5\baselineskip}
{
\center\normalsize\noindent\ignorespaces\@author\par
}%
\vspace{0.5\baselineskip}
{
\center\normalsize\ignorespaces\@institute\par
}%
\vspace{2\baselineskip}
}%

\let\OLDthebibliography\thebibliography%
\renewcommand\thebibliography[1]{%
\OLDthebibliography{#1}%
\setlength{\parskip}{0pt}%
\setlength{\itemsep}{0pt plus 0.3ex}%
}%

\begin{document}
\title{Algebraic structures in $\kappa$-Poincar\'e invariant gauge theories}
\author{Kilian Hersent$^a$, Philippe Mathieu$^b$, Jean-Christophe Wallet$^a$}
\institute{%
\textit{$^a$IJCLab, CNRS, 
University Paris-Saclay, 91405 Orsay, France\\
$^b$Department of Mathematics, 
University of Notre Dame\\Notre Dame, IN46556, USA
}\\%
e-mail: {\texttt{kilian.hersent@universite-paris-saclay.fr}}, {\texttt{pmathieu@nd.edu}}, {\texttt{jean-christophe.wallet@universite-paris-saclay.fr}}\\[1ex]%
}%
\maketitle

\begin{abstract}
$\kappa$-Poincar\'e invariant gauge theories on $\kappa$-Minkowski space-time, which are noncommutative analogs of the usual $U(1)$ gauge theory, exist only in five dimensions. These are built from noncommutative twisted connections on a hermitian right module over the algebra coding the $\kappa$-Minkowski space-time. We show that twisting the action of this algebra on the hermitian module, assumed to be a copy of it, affects neither the value of the above dimension nor the noncommutative gauge group defined as the unitary automorphisms of the module leaving the hermitian structure unchanged. Only the hermiticity condition obeyed by the gauge potential becomes twisted. Similarities between the present framework and algebraic features of twisted spectral triples are exhibited.
\end{abstract}
\newpage

\section{Introduction}\label{section1}

It is rather widely believed or even expected that some noncommutative structures manifest themselves near the Planck scale \cite{dopplic} when Quantum Gravity effects cease to be negligible \cite{amelin}, which would sign the unadequacy of the standard description of the space-time as a smooth manifold in any attempt to reconcile gravity and quantum mechanics. Many noncommutative (quantum) spaces have been considered in the literature for almost three decades but, so far, the most physically promising quantum space-time is the $\kappa$-Minkowski space \cite{luk2}. \\
This latter space appeared soon after the birth of the $\kappa$-Poincar\'e algebra \cite{luk1}, whose aim was to provide a deformation of the relativistic symmetries by using the framework of Hopf algebras and quantum groups \cite{leningrad}. This introduces a natural mass scale which is observer independent, represented by the deformation parameter $\kappa$ with mass dimension $1$, usually identified with the Planck mass in the 4-dimensional developments. Such a deformation of the Poincar\'e symmetry implies in general the appearance of deformed kinematics and provides a typical realization of the Doubly Special Relativity (DSR) \cite{ame-ca1} and \cite{AC1} suggested as a plausible modelisation of a ``flat-limit'' of quantum gravity. \\

As shown in \cite{majid-ruegg},  the $\kappa$-Poincar\'e Hopf algebra $\mathcal{P}^d_\kappa$  is characterized by a bicrossed-product structure. This structure basically shows that its rotations and boosts sub-algebra acts on the subalgebra generated by the so-called ``deformed translations''. But this latter has a back-action on the former and moreover is the dual algebra of the algebra modeling the $\kappa$-Minkowski space, on which also acts by duality the rotation and boost of the $\kappa$-Poincar\'e algebra. This exhibits clearly the rigid link existing between the $\kappa$-Poincar\'e algebra and the $\kappa$-Minkowski space, with a natural interpretation of the $\kappa$-Poincar\'e algebra as modeling the quantum space-time symmetries of the $\kappa$-Minkowski space.\\

These physically appealing features have generated a huge literature, mainly focused on related algebraic properties and their physical implications \cite{amelin} or on the exploration of classical properties of Noncommutative Field Theories (NCFT) on $\kappa$-Minkowski spaces 
\cite{habsb-impbis}-\cite{hrvat-1}. However, their quantum properties stayed poorly explored until recently \cite{mercati1}, \cite{PW2018} and \cite{PW2018bis}. This situation contrasted with the amount of literature dealing with perturbative quantum and renormalisability properties of NCFT on other noncommutative spaces, such as Moyal spaces $\mathbb{R}^{2n}_\theta$ \cite{Grosse-1}-\cite{vign-sym} or $\mathbb{R}^3_\lambda$ \cite{wal-16}, even including the more delicate case of gauge theories. \\

As shown in \cite{PW2018}, the use of a convenient star-product defining the $\kappa$-deformation of the Minkowski space permits one to investigate easily the perturbative properties of NCFT on $\kappa$-Minkowski spaces, thus realizing an actual progress in the area of these NCFT. This star-product together with its associated involution, a deformation of the usual hermitian conjugation, are given by 
\begin{align}
(f\star g)(x)&=\int \frac{dp^0}{2\pi}dy_0\ e^{-iy_0p^0}f(x_0+y_0,\vec{x})g(x_0,e^{-p^0/\kappa}\vec{x})  \label{starpro-4d},\\
f^\dag(x)&= \int \frac{dp^0}{2\pi}dy_0\ e^{-iy_0p^0}{\bar{f}}(x_0+y_0,e^{-p^0/\kappa}\vec{x})\label{invol-4d},
\end{align}
for any $f,g$ in a suitable multiplier algebra, say $\mathbb{A}$,  nicely characterized in \cite{DS} as the algebra of smooth functions with polynomial bounds together with all their derivatives and such that their inverse Fourier transform w.r.t. the $x_0$ variable is compactly supported. The associative $*$-algebra thus modeling the $\kappa$-Minkowski space is therefore defined by $\mathcal{M}^d_\kappa:=(\mathbb{A},\star,\dag)$. We will denote the $\kappa$-Poincar\'e algebra by $\mathcal{P}^d_\kappa$. Its essential algebraic properties together with those defining $\mathcal{M}^d_\kappa$ as a left module over $\mathcal{P}^d_\kappa$ are collected in the appendix. The superscript $d$ indicates the dimension of the $\kappa$-Minkowski space. We will denote the subalgebra of $\mathcal{P}^d_\kappa$ of deformed translation by $\mathcal{T}_\kappa^d$.\\

Recall that the star-product \eqref{starpro-4d} is obtained from a mere extension of the old construction of von Neumann \cite{old1} formalizing the Weyl quantization of the phase space\cite{old2}. This construction, which is very natural when the noncommutativity is of ``Lie algebra type'', basically exploits the main properties of the convolution algebra{\footnote{$\circ$ denotes the convolution product.}} $\mathbb{C}[\mathcal{G}]=(L^1(\mathcal{G}),\circ)$ for the group $\mathcal{G}$ linked to a given coordinates algebra. It has already been proved useful in the case of the quantum space $\mathbb{R}^3_\lambda$ for which the Lie algebra of coordinates is $\mathfrak{su}(2)$, leading to exploit properties of the convolution algebra of $SU(2)$\cite{poulinerie}. In the present situation, it combines the Weyl-Wigner quantization map with the convolution algebra of the affine group $\mathbb{R}\ltimes\mathbb{R}^{(d-1)}$ which is related to the Lie algebra of coordinates given by 
\begin{equation}
[x_0,x_i]=\frac{i}{\kappa}x_i,\ \ [x_i,x_j]=0,\ \ i,j=1,\cdots, (d-1),\label{simplist}
\end{equation}
where $x_0,\ x_i$, are the so-called ``noncommutative coordinates'' and $\kappa>0$ is the deformation parameter. This Lie algebra of coordinates is the simplest presentation of the noncommutativity ruling the $\kappa$-Minkowski space.\\

As far as gauge theories on $\kappa$-Minkowski space are concerned \cite{dimitriev} and \cite{rev-gauge}, requiring that the corresponding actions should be $\kappa$-Poincar\'e invariant in addition to gauge invariance appears to be physically natural. Indeed, one simply observes that Poincar\'e invariance holds in any acceptable field theory so that the whole scheme should be affected by the deformation. Putting all together, it is thus natural to look for  noncommutative gauge theories on $\mathcal{M}^d_\kappa$ described by polynomial actions depending on the curvature of the noncommutative connection satisfying the following two assumptions:
\begin{enumerate}[1)]
\item The action, say $S_\kappa$, is both invariant under $\mathcal{P}_\kappa^d$ and the  noncommutative $U(1)$ gauge symmetry, 
\item The commutative limit of $S_\kappa$, i.e. the limit $\kappa\to\infty$, coincides with the action describing an ordinary field theory. 
\end{enumerate}
As pointed out in \cite{PW2018}, the $\kappa$-Poincar\'e invariance of $S_\kappa$ is obtained for any action of the form $S_\kappa=\int\ d^dx\mathcal{L}$, i.e. when the trace involved in the action is the simple Lebesgue integral. This latter however is no longer cyclic with respect to the star-product \eqref{starpro-4d}. This makes difficult to reconcile gauge invariance with $\kappa$-Poincar\'e invariance. Indeed, a twist, the modular twist, depending on the dimension $d$ of $\mathcal{M}^d_\kappa$, appears upon cyclic permutation of the factors inside the trace, thus preventing the various factors arising from gauge transformations to balance each other\footnote{Recall that the situation is different in the case of $\mathbb{R}^4_\theta$ Moyal spaces for which the natural trace, still defined by the usual Lebesgue measure, is cyclic w.r.t. the Moyal star-product which thus does not require to introduce twists to insure gauge invariance.}. In \cite{MW20}, it has been shown that $\kappa$-Poincar\'e invariant gauge theories on $\mathcal{M}^d_\kappa$ satisfying the above two assumptions must be necessarily 5-dimensional. In particular, gauge invariance is achieved thanks to the existence of a unique twisted noncommutative differential calculus based on a unique family of twisted derivations of the algebra of the deformed translations $\mathcal{T}^d_\kappa$. Hence, as a physical outcome, the existence of an extra dimension is predicted. Some physical consequences arising from the $4$-dimensional effective theory stemming from simple compactification scenarii have been studied in \cite{MW20-bis} while, taking advantage of the BRST framework linked to the twisted gauge symmetry \cite{brst}, a first exploration of one-loop perturbative properties has been carried out in \cite{HMW1}, evidencing a kind a radiative breaking of the noncommutative gauge invariance through a vacuum instability against quantum fluctuations. \\

The above conclusions, in particular the special value d=5, hold true for a noncommutative analog of a $U(1)$ gauge theory. This latter is based on the notion of noncommutative connection on a hermitian right module, denoted by $\mathbb{E}$, assumed in \cite{MW20} and \cite{HMW1} to be a copy of $\mathcal{M}^d_\kappa$, acted on by the algebra via a right ``star-multiplication''. But this action could be twisted. Thus, a natural question is to determine in what extend the introduction of these additional twists may alter the above conclusions, and more specifically if such twists could make it possible to obtain $d=4$, which is a much more satisfying result from a phenomenological viewpoint. This is the purpose of the present paper.\\
We find that the hermiticity condition affecting the connection 1-form (gauge potential) can become twisted, thus departing from the usual condition $A^\dag=A$. But neither the noncommutative gauge group, as unitary elements of the module preserving the hermitian structure, nor the above special value d=5 at which gauge invariance and $\kappa$-Poincar\'e invariance fit together is altered by these additional twists.\\
In Section \ref{section2}, we present the noncommutative twisted differential geometrical framework. Section \ref{section3} deals with twisted connections, curvatures and twisted gauge transformations for the possible twisted actions of the algebra  $\mathcal{M}^d_\kappa$ on $\mathbb{E}$. In Section \ref{section4}, we discuss the results; in particular, we present the link existing between the present noncommutative twisted geometrical framework and the one of the Twisted Spectral Triples introduced in \cite{como-1}, whose particular forms have been used recently in the context of extended version \cite{martinet} of the Standard Model {\it{``\`a la Connes''}} \cite{alaconnes}, see in particular \cite{Devastato-Martinetti, Filaci-Martinetti}. Some phenomenological consequences are also presented. In Section \ref{section5}, we conclude.

\paragraph{Notations:}
Latin indices are strictly positive and refer to space-like coordinates, while the $0$ index refers to time-like coordinates. Greek indices are positive or zero and refer to space-time coordinates. We also make use of Einstein summation convention (assuming a Euclidean metric) unless otherwise stated. Moreover, we write $x:=(x_\mu)=(x_0,\vec{x})$ and $x\cdot y:= x_\mu y^\mu=x_\mu y_\nu\delta^{\mu\nu} = x_0y_0+\vec{x}\cdot\vec{y}$. The Fourier transform of $f\in L^1(\mathbb{R}^d)$ is 
\begin{equation}
(\mathcal{F}f)(p):=\int d^dx\ e^{-i(p_0x_0+\vec{p}\cdot\vec{x})}f(x)
\end{equation}
and $\bar{f}$ is its complex conjugate. $\mathcal{S}_c$ is the space of Schwartz functions with compact support in the first variable.\\

\section{Twisting differential calculs, connections and curvatures.}\label{section2}

\subsection{Twisted differential calculus.}\label{section21}

Let $\mathcal{D}_\gamma$, $\gamma\in\mathbb{R}$, denote the unique Lie abelian algebra of twisted derivations \cite{MW20} of the Hopf subalgebra of deformed translations $\mathcal{T}^d_\kappa\subset\mathcal{P}^d_\kappa$. The action of $\mathcal{T}_\kappa^d$ on $\mathcal{M}_\kappa^d$ is given by 
\begin{align}
(\mathcal{E}\triangleright f)(x) &= (e^{-P_0/\kappa}\triangleright f)(x) = f(x_0+\frac{i}{\kappa},\vec{x})\\
(P_\mu\triangleright f)(x) &= -i(\partial_\mu f)(x).
\end{align}
The algebra $\mathcal{D}_\gamma$ is defined by
\begin{equation}
\mathfrak{D}_\gamma=\big\{X_\mu:\mathcal{M}^d_\kappa\to \mathcal{M}^d_\kappa,\ \  X_0=\kappa\mathcal{E}^\gamma(1-\mathcal{E}),\ \ X_i=\mathcal{E}^\gamma P_i,\ \  i=1,2,...,d-1\big\}\label{tausig-famil}
\end{equation}
with $[X_\mu,X_\nu]:=X_\mu X_\nu-X_\nu X_\mu=0$ and verifying the twisted Leibniz rule
\begin{equation}
X_\mu(f\star h)=X_\mu(f)\star \mathcal{E}^\gamma(h)+ \mathcal{E}^{1+\gamma}(f)\star X_\mu(h),\label{tausigleibniz}
\end{equation}
for any $f,h\in\mathcal{M}_\kappa^d$, which is consistent with the coproduct $\Delta:\mathcal{P}^d_\kappa\to\mathcal{P}^d_\kappa$ equipping $\mathcal{P}^d_\kappa$, whose action on the primitive elements of $\mathcal{T}^d_\kappa$, $(\mathcal{E},\ P_\mu)$, is given by 
\begin{align}
\Delta P_0 &= P_0\otimes\mathbb{I}+\mathbb{I}\otimes P_0,\\
\Delta P_i &= P_i\otimes\mathbb{I}+\mathcal{E}\otimes P_i,\\
\Delta \mathcal{E} &=\mathcal{E}\otimes\mathcal{E}.
\end{align}

Recall that $\mathfrak{D}_\gamma$ is a $\mathcal{Z}(\mathcal{M}_\kappa^d)$-bimodule, since one easily verifies that $(X.z)(f):=X(f)\star z=z\star X(f)=(z\cdot X)(f)$, for any $f\in\mathcal{M}_\kappa^d$ and any $z\in \mathcal{Z}(\mathcal{M}_\kappa^d)$, the center of $\mathcal{M}_\kappa^d$.\\

Recall that the differential calculus based on the Lie algebra of twisted derivations $\mathfrak{D}_\gamma$ is defined \cite{MW20} by the following differential algebra
\begin{equation}
\left(\Omega^\bullet=\bigoplus_{n=0}^{d}\Omega^n(\mathfrak{D}_\gamma),\times,{\bf{d}}\right),\label{diff-algebra}
\end{equation}
for any $\omega\in\Omega^m(\mathfrak{D}_\gamma)$, $\eta\in\Omega^n(\mathfrak{D}_\gamma)$, where $\Omega^n(\mathfrak{D}_\gamma)$ is the linear space of $n$-linear{\footnote{Here, the linearity holds w.r.t. $\mathcal{Z}(\mathcal{M}_\kappa^d)$,  the center of the algebra.}} antisymmetric forms, $\alpha:\mathfrak{D}^{n}_\gamma\to\mathcal{M}_\kappa^d$ such that $\alpha(X_1,X_2,...,X_n)\in\mathcal{M}_\kappa^d$ and 
\begin{equation}
\alpha(X_1,X_2,...,X_n\cdot z)=\alpha(X_1,X_2,...,X_n)\star z,
\end{equation}
for any $z\in\mathcal{Z}(\mathcal{M}_\kappa^d)$ and any $X_1, \hdots, X_n\in\mathfrak{D}_\gamma$. Note that $\Omega^0(\mathfrak{D}_\gamma)=\mathcal{M}_\kappa^d$.\\

In \eqref{diff-algebra}, the associative product $\times:\Omega^m(\mathfrak{D}_\gamma)\otimes\Omega^n(\mathfrak{D}_\gamma)\to\Omega^{m+n}(\mathfrak{D}_\gamma)$  is defined by
\begin{align}
\nonumber
&(\omega\times\eta)(X_1,...,X_{m+n})\\
&\qquad\qquad=\frac{1}{m!n!}\sum_{s\in\mathfrak{S}(m+n)}(-1)^{\text{sign}(s)}\omega(X_{s(1)},...,X_{s(m)})\star \eta(X_{s(m+1)},...,X_{s(n)})\label{ncwedge1},
\end{align}
where $\mathfrak{S}(m+n)$ denotes, as usual, the symmetric group of a set of $m+n$ elements and $\text{sign}(s)$ is the signature of the permutation $s$. 
The differential ${\bf{d}}:\Omega^m(\mathfrak{D}_\gamma)\to\Omega^{m+1}(\mathfrak{D}_\gamma)$ for any $m={0},1,...,{\left(d-1\right)}$, satisfying ${\bf{d}}^2=0$, is defined by 
\begin{equation}
\left({\bf{d}}\omega\right)\left(X_1,X_2,...,X_{p+1}\right)
=\sum_{i=1}^{p+1}(-1)^{i+1}X_i\left(\omega(X_1,...,\vee_i,...,X_{p+1})\right), \label{ncwedge2}
\end{equation}
where the symbol $\vee_i$ means that the derivation $X_i$ is omitted. It obeys the following twisted Leibniz rule
\begin{equation}
{\bf{d}}(\omega\times\eta)={\bf{d}}\omega\times\mathcal{E}^\gamma(\eta)+(-1)^{\delta(\omega)}\mathcal{E}^{1+\gamma}(\omega)\times{\bf{d}}\eta\label{leibniz-form},
\end{equation}
where $\delta({\omega})$ is the degree of $\omega$ and $\mathcal{E}^x(\omega)\in\Omega^n(\mathfrak{D}_\gamma)$ is defined for any $x\in\mathbb{R}$ and any $\omega\in\Omega^n(\mathfrak{D}_\gamma)$ by
\begin{equation}
\mathcal{E}^{x}(\omega)(X_1,...X_n)=\mathcal{E}^x(\omega(X_1,...X_n))
\end{equation}
for any $X_1, \hdots, X_n\in\mathfrak{D}_\gamma$.

Note that the differential algebra \eqref{diff-algebra} is not graded commutative since one can easily verify that 
\begin{equation}
\omega\times\eta\ne(-1)^{\delta(\omega)\delta(\eta)}\eta\times\omega.
\end{equation}
Note also that the derivations are {\it{not}} real derivations. Indeed, a simple computation yields 
\begin{equation}
(X_\mu(f))^\dag=-\mathcal{E}^{-2\gamma-1}(X_\mu(f^\dag))\label{xpasreel}
\end{equation}
which holds for any $f\in\mathcal{M}^d_\kappa$.

\subsection{Twisting connections and curvatures - Assumptions.}\label{section22}

There are three technical assumptions supplementing the basic structural assumptions listed at the end of Section \ref{section1} which will underly the ensuing analysis:
\begin{enumerate}[1)]
\item We assume that the twisted versions of noncommutative connection are defined on a right module $\mathbb{E}$ over $\mathcal{M}^d_\kappa$, which is suitable for our purpose of constructing noncommutative analogs of the (flat) Yang-Mills theory.
\item The module $\mathbb{E}$  is assumed to be a copy of $\mathcal{M}^d_\kappa$, i.e. $\mathbb{E}\simeq\mathcal{M}^d_\kappa$. This somewhat simplifying assumption allows one to model noncommutative analog of the $U(1)$ gauge symmetry\footnote{Generalisation to an analog of U(n) can be obtained from a module built from the product of n copies of $\mathbb{E}$.}.
\item We assume that the action of $\mathcal{M}^d_\kappa$ on $\mathbb{E}$, defined as a linear map $\Phi:\mathbb{E}\otimes\mathcal{M}^d_\kappa\to\mathbb{E}$, is twisted by an automorphim of $\mathcal{M}^d_\kappa$. Namely, the action is assumed to be 
\begin{equation}
\Phi(m\otimes f):=m\bullet f=m\star \sigma(f),\ \ \sigma\in\text{Aut}(\mathcal{M}^d_\kappa)\label{zeaction},
\end{equation}
for any $m\in\mathbb{E}$, $f\in\mathcal{M}^d_\kappa$.
\end{enumerate}

At this point, two comments are in order:\\
\begin{itemize}
\item We note that this last assumption extends the analysis carried out in \cite{MW20} where only an untwisted action is assumed, instead of \eqref{zeaction}. One may wonder if the appearance of such an additional twist $\sigma$ may modify (when suitably chosen) the value of the dimension of $\mathcal{M}^d_\kappa$ for which both gauge invariance and $\kappa$-Poincar\'e invariance can be achieved. It turns out that this value does not depend on $\sigma$ as we will see in a while.\\
\item Recall that \eqref{zeaction} satisfies the relations
\begin{equation}
(m\bullet f)\bullet h=m\bullet(f\star h),\ \ m\bullet 1=m\label{relations}.
\end{equation}
for any $f\in\mathcal{M}^d_\kappa$, $m\in\mathbb{E}$. It follows that any twisted action of the form 
\begin{equation}
m\bullet f=\theta(m)\star f
\end{equation}
for any automorphism $\theta$ of the module $\mathbb{E}$, would satisfy the second relations of \eqref{relations} if and only if $\theta\left(m\right) = m$ for any $m\in\mathbb{E}$. Equivalently, $\theta=\text{Id}$, which thus would correspond to the case considered in \cite{MW20}. In short, \eqref{zeaction} is the only nontrivial twisted action of the algebra on the right module.
\end{itemize}
Having in mind to build a physically relevant model, it turns out that the module $\mathbb{E}$ must be promoted to the status of hermitian module. Thus, $\mathbb{E}$ must be equipped with a hermitian structure, i.e. a sesquilinear map 
$h: \mathbb{E}\times\mathbb{E}\to \mathcal{M}^d_\kappa$ such that
\begin{eqnarray}
 h(m\bullet f,n\bullet k)&=&f^\dag \star h(m,n)\star k,\label{gene-structerm}\\
h(m,n)^\dag&=&h(n,m),\label{chermitik}
\end{eqnarray}
for any $m,n\in\mathbb{E}$, $f,k\in\mathcal{M}^d_\kappa$, supplemented with $h(1,1)=1$. From \eqref{zeaction} and \eqref{gene-structerm}, one infers
$h(1\bullet f,1\bullet k)=f^\dag \star h(1,1)\star k=f^\dag\star k$ while one also has $h(1\bullet f,1\bullet k)=h(\sigma(f),\sigma(k))$. It follows that 
\begin{equation}
h(f,k)=(\sigma^{-1}(f))^\dag\star \sigma^{-1}(k)\label{hermit}
\end{equation}
for any $f,k\in\mathcal{M}^d_\kappa\simeq\mathbb{E}$.\\

The condition \eqref{gene-structerm} expresses the compatibility between the hermitian structure and the action of the algebra on the module defined by \eqref{zeaction}. Equations \eqref{gene-structerm} and \eqref{chermitik} are the main conditions entering the standard definition of a hermitian structure. This is the one we will use in the next section.

\section{Twisted connections from hermitian structure.}\label{section3}

\subsection{Connection, curvature and gauge transformations.}\label{section31}
Given the hermitian structure \eqref{hermit}, the unitary gauge transformations are defined as the automorphisms of the module $\mathbb{E}$ preserving \eqref{hermit}. \\

For any
$\varphi\in\text{Aut}(\mathbb{E})$, it follows, for any $m\in\mathbb{E}$, $f\in\mathcal{M}^d_\kappa$, that $\varphi(m\bullet f)=\varphi(m)\bullet f=\varphi(m)\star\sigma(f)$. This, combined with $1\bullet f=1\star\sigma(f)=\sigma(f)$ and setting $\varphi(1)=g\in\mathbb{E}$, yields $\varphi(\sigma(f))=g\star\sigma(f)$ and thus $\varphi(f)=g\star f$. Then, setting
\begin{equation}
f^g:=g\star f\label{fg},
\end{equation}
one computes  $h(f^g,k^g)=\sigma^{-1}(f)^\dag\star \sigma^{-1}(g)^\dag\star \sigma^{-1}(g)\star \sigma^{-1}(k)=h(f,k)$ where the rightmost equality holds provided $\sigma^{-1}(g)^\dag\star \sigma^{-1}(g)=1$. Hence, the unitary gauge group is given by
\begin{equation}
U_\kappa^\sigma(1):=\{g\in\mathbb{E}\simeq\mathcal{M}^d_\kappa,\ \  \sigma^{-1}(g)^\dag\star \sigma^{-1}(g)=1\}\label{unitar-transf}.
\end{equation}
We define the unitary group, the noncommutative analog of the $U(1)$ group, as
\begin{equation}
U_\kappa=U_\kappa^{\sigma=\text{Id}}(1)=\{g\in\mathbb{E}\simeq\mathcal{M}^d_\kappa,\ \  g^\dag\star g=1\}\label{unitar-transf-triv}.
\end{equation}
For the moment, we do not assume additional properties for the automorphism $\sigma$. Specific implication for $\sigma$ to be either a $*$-automorphism or a regular automorphism \cite{como-1} will be analysed in a while.\\

To characterize the twisted connection, one looks for a map $\nabla:\mathbb{E} \times\mathfrak{D}_\gamma\to\mathbb{E}$ defined for any $m\in\mathbb{E}$, $f\in\mathcal{M}^d_\kappa$, $X_\mu\in\mathfrak{D}_\gamma$ by
\begin{eqnarray} 
\nabla_{X_\mu}(m\bullet f) &= &\nabla_{X_\mu}(m)\bullet\tau_{1,\mu}(f) + \tau_{2,\mu}(m)\bullet X_\mu(f),\label{gene-conn}
\end{eqnarray}
where $\tau_{1,\mu}$ and $\tau_{2,\mu}$ are some twists and no Einstein summation over the repeated indices $\mu$ is understood, which is supplemented by
\begin{eqnarray} 
\nabla_{X_\mu+X^\prime_\mu}(m)&=&\nabla_{X_\mu}(m)+\nabla_{X^\prime_\mu}(m)\label{enplus1}\\
\nabla_{z\cdot X_\mu}(m)&=&\nabla_{X_\mu}(m)\star z\label{enplus2}.
\end{eqnarray}
for any $m\in\mathbb{E}$, $X_\mu, X'_\mu\in\mathfrak{D}_\gamma$ and $z\in\mathcal{Z}(\mathcal{M}^d_\kappa)$.

We set
\begin{equation}
\nabla_\mu:=\nabla_{X_\mu}.
\end{equation}
In \eqref{gene-conn}, the twists $\tau_{1,\mu}$ and $\tau_{2,\mu}$ can be entirely determined by a combination property related to  \eqref{tausigleibniz} and \eqref{relations}, as we now show.\\

By computing both sides of the identity $\nabla_\mu(m\bullet(f\star k)) = \nabla_\mu((m\bullet f)\bullet k)$, one easily obtains $\tau_{1,\mu}(k)=\mathcal{E}^\gamma(k)$ for any $k\in\mathcal{M}^d_\kappa$, and thus
\begin{equation}
 \tau_{1,\mu}=\mathcal{E}^\gamma\label{equation1},
\end{equation}
while a second relation takes the form $\tau_{2,\mu}(m\bullet f)\bullet X_\mu(k)= \tau_{2,\mu}(m)\bullet\mathcal{E}^{\gamma+1}(f)\bullet X_\mu(k)$ (recall no summation over the indices $\mu$) so that
\begin{equation}
\tau_{2,\mu}(m\bullet f)= \tau_{2,\mu}(m)\bullet\mathcal{E}^{\gamma+1}(f)\label{relation2},
\end{equation}
 for any $f\in\mathcal{M}^d_\kappa$, $m\in\mathbb{E}$ which, assuming $\tau_{2,\mu}(1)=1$, is verified provided $[\sigma,\mathcal{E}]=0$ and
\begin{equation}
\tau_{2,\mu}=\mathcal{E}^{\gamma+1}\label{equation2}.
\end{equation}
Hence $\tau_{1,\mu}$ and $\tau_{2,\mu}$ do not depend on $\mu$. Notice that this independence can be shown by using the $\mathcal{Z}(\mathcal{M}_\kappa^{d})$-linearity of $\nabla_\mu$ and writing, for any $z\in\mathcal{Z}(\mathcal{M}_\kappa^{d})$, $\nabla_{X_\mu+z\cdot X_\nu}(m\bullet f)$ in two different ways.\\

By combining \eqref{gene-conn} with \eqref{equation1} and\eqref{equation2} and setting $m=1$, one easily obtains 
\begin{equation}
\nabla_{\mu}(\sigma( f) )= \nabla_{\mu}(1)\star\mathcal{E}^\gamma(\sigma(f) )+ X_\mu(\sigma(f)),\label{truffe}
\end{equation}
for any $f\in\mathcal{M}^d_\kappa$ and therefore
\begin{equation}
\nabla_{\mu}(f)= A_\mu\star\mathcal{E}^\gamma(f) + X_\mu(f)\label{dercov},
\end{equation}
where we have set
\begin{equation}
A_\mu:=\nabla_{\mu}(1).\label{potential}
\end{equation}
Notice that the twisted connection does not depend on the twist $\sigma$ affecting the action of the algebra on the module. Notice also in \eqref{truffe} the appearance of $\sigma(f)$ as the argument instead of $f$, hence illustrating the last statement in Comment 2) given in Subsection \ref{section22}.\\

As expected, the hermiticity condition obeyed by the connection is twisted, as it was already the case for the untwisted action of $\mathcal{M}^d_\kappa$ on $\mathbb{E}$ \cite{MW20}. Algebraic manipulations, using in particular the fact that $X_\mu$ is not a real derivation as given by \eqref{xpasreel}, yield
\begin{equation}
h(i\mathcal{E}^{-2\gamma-1}(\nabla_{X_{\mu}}(m_1)),\mathcal{E}^{\gamma}(m_2)+h(\mathcal{E}^{-\gamma-1} (m_1),i\nabla_{X_{\mu}}(m_2))=iX_\mu h(m_1,m_2)\label{cledag}
\end{equation}
for any $X_\mu\in\mathfrak{D}_\gamma$, $m_1,m_2\in\mathcal{M}_\kappa^d$, which holds true provided
\begin{equation}
\sigma^{-1}(A_\mu)=\mathcal{E}^{2\gamma+1}(\sigma^{-1}(A_\mu)^\dag)\label{twistedhermit},
\end{equation}
as it can be easily obtained by simple algebraic manipulations. Note that \eqref{twistedhermit} coincides with the twisted hermiticity condition for $A_\mu$ found in \cite{HMW1} when $\sigma=\text{Id}$.\\

At this point, one important point must be outlined. It turns out that some (but not all) the conclusions that will come out depend whether the automorphism $\sigma$ is a usual $*$-automorphism, i.e. such that $\sigma(f)^\dag=\sigma(f^\dag)$ for any $f\in\mathcal{M}^d_\kappa$ or is instead a regular automorphism \cite{como-1}, i.e. verifying $\sigma(f)^\dag=\sigma^{-1}(f^\dag)$ for any $f\in\mathcal{M}^d_\kappa$. We will come back to these regular automorphisms in Subsection \ref{section41}. We now examine successively these two possibilities.\\

\subsection{$\sigma$ as a $*$-automorphism.}\label{section32}

Assume that $\sigma$ is a $*$-automorphism; i.e. one has $\sigma(f)^\dag=\sigma(f^\dag)$.  Then, the defining relation of the gauge group \eqref{unitar-transf} reads $\sigma^{-1}(g^\dag)\star \sigma^{-1}(g)=\sigma^{-1}(g^\dag\star g)=1$ which is equivalent to $g^\dag\star g=1$. It follows that the gauge group $U_\kappa^\sigma(1)$ reduces to the unitary gauge group arising when the action of the algebra $\mathcal{M}^d_\kappa$ on the module $\mathbb{E}$ is not twisted. This latter group is denoted by $\mathcal{U}$ in \cite{MW20}. Hence, 
\begin{equation}
U_\kappa^\sigma(1)=U_\kappa(1),
\end{equation}
where $U_\kappa$ is defined in \eqref{unitar-transf-triv} and the gauge group does not depend on the twist $\sigma$.\\
Besides, the hermiticity condition \eqref{twistedhermit} boilts down to
\begin{equation}
A_\mu=\mathcal{E}^{2\gamma+1}(A_\mu)^\dag\label{twist-ancien},
\end{equation}
which, as expected, is independent of $\sigma$ and reproduces the hermiticity condition of \cite{HMW1}.\\

Now, it can be easily realized that the analysis of \cite{MW20} can be thoroughly reproduced. Indeed, define the gauge transformation of the connection as
\begin{equation}
\nabla_\mu^g(.)=\rho_1(g^\dag)\star\nabla_\mu(\rho_2(g)\star.)\label{gauge-connec-gene},
\end{equation}
where $\rho_1$ and $\rho_2$ are two regular automorphisms of $\mathbb{E}$. Equation \eqref{gauge-connec-gene} still defines a connection provided
\begin{equation}
\rho_1(g^\dag)\star\mathcal{E}^{\gamma+1}\rho_2(g)=1\label{connec-stable},
\end{equation}
for any $g\in U_\kappa(1)$. Then, the curvature $\mathcal{F}_{\mu\nu}:=\mathcal{F}(X_\mu,X_\nu):\mathbb{E}\to\mathbb{E}$, defined as
\begin{equation}
\mathcal{F}_{\mu\nu}=\mathcal{E}^{1-\gamma}(\nabla_\mu\mathcal{E}^{-1-\gamma}\nabla_\nu-\nabla_\nu\mathcal{E}^{-1-\gamma}\nabla_\mu)\label{decadix},
\end{equation}
is a morphism of module, i.e. 
\begin{equation}
\mathcal{F}_{\mu\nu}(m\bullet f)=\mathcal{F}_{\mu\nu}(m)\bullet f\label{morphimod},
\end{equation}
for any $m\in\mathbb{E}$, $f\in\mathcal{M}^d_\kappa$.

On the other hand, the field strength defined as
\begin{equation}
F_{\mu\nu}:=\mathcal{F}_{\mu\nu}(1),
\end{equation}
transforms covariantly as
\begin{equation}
F_{\mu\nu}^g=\mathcal{E}^{1-\gamma}\rho_1(g^\dag)\star{F}_{\mu\nu}\star \rho_2(g),\label{decadix1}
\end{equation}
 if
\begin{equation}
\rho_2(g)\star \mathcal{E}^{-1- \gamma}\rho_1(g^\dag)=1\label{rel3},
\end{equation}
for any $g\in U_\kappa(1)$.\\

The combination of \eqref{connec-stable}, \eqref{rel3} and the unitarity condition defining $U(1)_\kappa$ \eqref{unitar-transf-triv} yields 
\begin{equation}
\rho_1=\mathcal{E}^{\gamma+1}\rho_2\label{cabouge}
\end{equation}
and
\begin{equation}
F_{\mu\nu}=\mathcal{E}^{-2\gamma}\triangleright(X_\mu A_\nu-X_\nu A_\mu)+(\mathcal{E}^{1-\gamma}\triangleright A_\mu)\star(\mathcal{E}^{-\gamma}\triangleright A_\nu)-(\mathcal{E}^{1-\gamma}\triangleright A_\nu)\star(\mathcal{E}^{-\gamma}\triangleright A_\mu) \label{efmunu},
\end{equation}
with the following gauge transformation
\begin{equation}
F_{\mu\nu}^g=\mathcal{E}^{2}\rho_2(g^\dag)\star{F}_{\mu\nu}\star \rho_2(g)\label{gaugetransf-courb},
\end{equation}
for any $g\in U_\kappa(1)$, which does not depend on $\gamma$. From \eqref{gauge-connec-gene}, one obtains the gauge transformations for the gauge potential given by
\begin{equation}
A_\mu^g=\mathcal{E}^{\gamma+1}\rho_2(g^\dag)\star A_\mu\star \mathcal{E}^\gamma\rho_2(g)+\mathcal{E}^{\gamma+1}\rho_2(g^\dag)\star X_\mu(\rho_2(g)).
\end{equation}
\\
The extension of the map \eqref{decadix} to a map $F:\mathbb{E}\to\mathbb{E}\otimes\Omega^2(\mathfrak{D}_\gamma)$ is straightforward. One obtains
\begin{equation}
F=\mathcal{E}^{-2\gamma}\triangleright {\bf{d}}A+\mathcal{E}^{-\gamma}\triangleright((\mathcal{E}\triangleright A)\times A)\label{curvat-f},
\end{equation}
thus introducing the curvature 2-form $F$, where $A$ represents the connection 1-form defined from \eqref{dercov} and \eqref{potential} by $\nabla:\mathbb{E}\to\mathbb{E}\otimes\Omega^1(\mathfrak{D}_\gamma)$, with
\begin{equation}
\nabla(f)=A\star f+ 1\otimes {\bf{d}}f\label{forms-alph1},
\end{equation}
for any $f\in\mathcal{M}^d_\kappa$. From a simple calculation, one can check that $F$ \eqref{curvat-f} satisfies the following Bianchi identity
\begin{equation}
{\bf{d}}F=(\mathcal{E}^{1+\gamma}\triangleright F)\times A-(\mathcal{E}^{2}\triangleright A)\times(\mathcal{E}^{\gamma}\triangleright F)\label{bianchi}.
\end{equation}

\subsection{$\sigma$ as a regular automorphism.}\label{section33}
Assume that $\sigma$ is a regular automorphism, i.e. it verifies
\begin{equation}
\sigma(f)^\dag=\sigma^{-1}(f^\dag)\label{regular1}
\end{equation}
for any $f\in\mathcal{M}^d_\kappa$. It follows that the defining relation of the gauge group \eqref{unitar-transf} now reads, for any $g\in\mathbb{E}$, 
\begin{equation}
 \sigma^{-1}(g)^\dag\star \sigma^{-1}(g)=\sigma(g^\dag)\star\sigma^{-1}( g)=u^\dag\star u=1,
\end{equation}
where $u$ is given by $u=\sigma^{-1}(g)$. Hence, the gauge group \eqref{unitar-transf} is isomorphic to $U_\kappa(1)$, i.e. one has 
\begin{equation}
U_\kappa^\sigma(1)\simeq\sigma(U_\kappa(1))\label{isomorgroup}.
\end{equation}
Besides, the hermiticity condition \eqref{twistedhermit} becomes
\begin{equation}
A_\mu=\mathcal{E}^{2\gamma+1}\sigma^2(A_\mu^\dag)\label{sigma-hermit}.
\end{equation}
Note that \eqref{sigma-hermit} reduces to the usual hermiticity condition $A_\mu^\dag=A_\mu$ when $\sigma^2=\mathcal{E}^{-2\gamma-1}$.\\

The analysis carried out in Subsection \ref{section32} can be easily adapted to the present situation. For that purpose, set:
\begin{equation}
\beta_1=\rho_1\sigma^{-1},\ \ \beta_2=\rho_2\sigma\label{celesbetas},
\end{equation}
in which $\beta_1$ and $\beta_2$ are regular automorphisms by construction. Then, the gauge transformations for the connection are defined by adapting \eqref{gauge-connec-gene}, giving rise to
\begin{equation}
\nabla_\mu^u(.)=\rho_1\sigma^{-1}(u^\dag)\star\nabla_\mu(\rho_2\sigma(u)\star.),\ \ \rho_1\sigma^{-1}(u^\dag)\star\mathcal{E}^{\gamma+1}\rho_2\sigma(u)=1,\label{zerelation}
\end{equation}
for any $u\in\mathbb{E}$ such that $u^\dag\star u=1$, i.e. any $u\in U_\kappa(1)$. The curvature, a morphism of module, is still given by \eqref{decadix} and the corresponding field strength $F_{\mu\nu}$ transforms as
\begin{equation}
F_{\mu\nu}^u=\rho_1\sigma^{-1}(u^\dag)\star F_{\mu\nu}\star\rho_2\sigma^{}(u)\label{newftransf}
\end{equation}
provided one has
\begin{equation}
\rho_2\sigma^{}(u)\star\mathcal{E}^{-1-\gamma}\rho_1\sigma^{-1}(u^\dag)=1,\label{relprim1}
\end{equation}
 for any $u\in U_\kappa(1)$. A simple inspection of \eqref{relprim1}, \eqref{newftransf} and the 2nd relation of \eqref{zerelation} shows that equation \eqref{cabouge} is changed into
\begin{equation}
\rho_1=\mathcal{E}^{\gamma+1}\sigma^2\rho_2\label{cabougeprime}.
\end{equation}
The gauge transformations take finally the form, for any $u\in U_\kappa(1)$:
\begin{equation}
F_{\mu\nu}^u=\mathcal{E}^{2}\rho_2\sigma(u^\dag)\star F_{\mu\nu}\star\rho_2\sigma^{}(u)\label{final-transf-F},
\end{equation}
and
\begin{equation}
A_\mu^g=\mathcal{E}^{\gamma+1}\rho_2\sigma^{-1}(u^\dag)\star A_\mu\star \mathcal{E}^\gamma\rho_2\sigma(u)+\mathcal{E}^{\gamma+1}\rho_2\sigma^{-1}(u^\dag)\star X_\mu(\rho_2\sigma(g)).
\end{equation}

\section{Discussion.}\label{section4}
\subsection{Linking to twisted spectral triples.}\label{section41}
Note that twisted differential calculi already appeared in the context of twisted Spectral Triples. These spectral triples have a relatively long story in mathematics, introduced in \cite{como-1} within the context of operator algebras of type III. They also occurred in relation with quantum groups for which twisted actions on algebras are natural. Twisted Spectral Triples have also been introduced recently as the master structures underlying refined versions \cite{martinet} of the Standard Model description {\it{``\`a la Connes''}} \cite{alaconnes}. We refer to \cite{Devastato-Martinetti, Filaci-Martinetti} for related detailed analysis. Notice that interesting versions of twisted spectral triples for $\kappa$-Minkowski spaces were considered
in \cite{Matassa2012, Matassa2013}, in which the corresponding Dirac operators are built from derivatives of $\mathcal{D}_\gamma$.\\

Recall that a twisted spectral triple must satisfy the condition that 
\begin{equation}
[D,f]_\rho:=Df-\rho(f)D\label{cnoixe}
\end{equation}
is such that $[D,f]_\rho\in\mathcal{B}(\mathcal{H})$, the space of bounded operators on some Hilbert space $\mathcal{H}$, for any $f$ in some suitable (involutive) algebra $\mathbb{A}$. Here, $D$ is the Dirac operator linked to the spectral triple, while $\rho\in\text{Aut}(\mathbb{A})$ is the twist which must be a regular automorphism, that is
\begin{equation}
\rho(f)^\dag=\rho^{-1}(f^\dag)\label{regular}
\end{equation}
for any $f\in\mathbb{A}$, a condition motivated by the requirement (for technical reasons) that $\mathbb{A}$ supports the action of a one-parameter group of $^*$-automorphisms $\rho_t$, $t\in\mathbb{R}$ such that $\rho_i$ is exactly the analytic extension of $\rho_t$. This group of automorphisms is called the modular group. This, of course, refers to the fascinating Tomita-Takesaki modular theory. For mathematical details, see \cite{takesaki}. \\
Coming back to the twisted spectral triples framework, it turns out that $[D,f]_\rho$ \eqref{cnoixe} acts as a twisted derivation on $\mathbb{A}$, namely (product of algebra understood)
\begin{equation}
\delta_\rho(fk):=[D,fk]_\rho=[D,f]_\rho k+\rho(f)[D,k]_\rho,
\end{equation}
for any $f,k\in\mathbb{A}$. Note that $\delta_\rho$ even extends to a derivation of $\mathbb{A}$ in the space of 1-forms $\Omega^1_D=\{\omega=\sum_i f_i[D,k_i],\ f_i,k_i\in\mathbb{A} \}$, provided $\Omega^1_D$ is a bimodule over $\mathbb{A}$ with the following action $f\bullet\omega\bullet k=\rho(f)\omega k$, for any $f,k\in\mathbb{A}$.\\

In the present situation, the group of $^*$-automorphisms of the algebra $\mathcal{M}_\kappa$ is given by
\begin{equation}
\rho_t(f)=e^{it\frac{P_0}{\kappa}}(f)=\mathcal{E}^{-it},\label{sigmat-modul}
\end{equation}
for any $t\in\mathbb{R}$ and $f\in\mathcal{M}^d_\kappa$. This is discussed at length in \cite{PW2018}. Recall in particular that the above group of automorphism is rigidly linked ot a KMS weight which is defined here by the map $\eta:\mathcal{M}^d_\kappa\to\mathbb{C}$, $\eta(f)=\int d^dx\ f$ and therefore linked to the twisted trace defined by the usual Lebesgue integral, as it can be expected from the modular theory.\\
From \eqref{sigmat-modul} and the above discussion, it follows that the role of the automorphism $\rho$ is defined by the action of $\mathcal{E}$, i.e.
\begin{equation}
\rho(f)=\mathcal{E}(f)
\end{equation}
for any $f\in\mathcal{M}^d_\kappa$. It can be easily verified that $\mathcal{E}$ acts as a regular automorphism, i.e.
\begin{equation}
(\mathcal{E}(f))^\dag=\mathcal{E}^{-1}(f^\dag),
\end{equation}
which is a mere consequence of \eqref{pairing-involution}.\\
Besides, the role of $\delta_\rho$ as a derivation on the algebra is simply played by the differential ${\bf{d}}:\Omega^0(\mathfrak{D}_0)\to\Omega^{1}(\mathfrak{D}_0)$ defined in \eqref{ncwedge2}, with however $\gamma=0$, and satisfying the twisted Leibniz rule \eqref{leibniz-form}.

\subsection{Phenomenological consequences.}\label{4.2}

From the analysis carried out in Section \ref{section3}, it appears that a twisted action of the algebra $\mathcal{M}^d_\kappa$ on the module $\mathbb{E}$ characterized by an automorphism $\sigma\in\text{Aut}(\mathcal{M}^d_\kappa)$ affects essentially the hermiticity condition ruling in particular the gauge potential $A_\mu$, as illustrated by \eqref{twist-ancien} and \eqref{sigma-hermit}, corresponding respectively to the case of $\sigma$ being a $*$-automorphism and a regular automorphism. However, the relevant ``noncommutative gauge group'' is essentially, up to an isomorphism, $U_\kappa(1)$, i.e. the noncommutative analog of $U(1)$, defined by \eqref{unitar-transf-triv}.\\

More interestingly, the above twist does not change the special value of the dimension $d$ of $\mathcal{M}^d_\kappa$ at which the $\kappa$-Poincar\'e invariant{\footnote{Recall that one has $h\blacktriangleright S=\int d^dx\ h\triangleright\mathcal{L}=\epsilon(h)S$ for any $h\in\mathcal{P}^d_\kappa$, where $\epsilon(.)$ is the co-unit of $\mathcal{P}^d_\kappa$, thus insuring the $\kappa$-Poincar\'e invariance of any action functional of the form $S=\int d^dx\ \mathcal{L}$.}}
real polynomial action 
\begin{equation}
S_\kappa=\int d^dx\ F_{\mu\nu}\star F_{\mu\nu}^\dag\label{zelagrangien}
\end{equation}
is $U_\kappa(1)$ gauge-invariant. To see that, consider for instance the gauge transformed curvature in the case of a $*$-automorphism \eqref{gaugetransf-courb}. We have
\begin{align}
\label{computation}
\int d^dx\ F_{\mu\nu}^g\star (F_{\mu\nu}^g)^\dag
&=\int d^dx\ \big({\mathcal{E}^{d-1-2}\rho_2(g)\star\mathcal{E}^2\rho_2(g^\dag)}\big)\star F_{\mu\nu}\star F_{\mu\nu}^\dag
\end{align}
where we used the twisted trace formula 
\begin{equation}
\int d^dx\ (f\star g)(x)=\int d^dx\ ((\mathcal{E}^{d-1}(g)\star f)(x). 
\end{equation}
The gauge invariance of the action reads
\begin{equation}
\int d^dx\ F_{\mu\nu}^g\star (F_{\mu\nu}^g)^\dag = \int d^dx\ F_{\mu\nu}\star F_{\mu\nu}^\dag
\end{equation}
which, according to \eqref{computation}, is true if and only if $\mathcal{E}^{d-1-2}\rho_2(g)\star\mathcal{E}^2\rho_2(g^\dag)=1$, which obviously occurs if and only if
\begin{equation}
d=5.
\end{equation}

Note that $g$ given just above is any element of the unitary gauge group $U_\kappa(1)$ therefore verifying the ``unitarity condition'' $g^\dag\star g=g\star g^\dag=1$, which is essential to fulfill the last equality in \eqref{computation}. A similar conclusion holds when $\sigma$ is regular.\\

One may wonder if twisting the compatibility condition defining the hermitian structure \eqref{gene-structerm} would alter this special value for $d$. It turns out that such a twist would not modify this result. Indeed, trade \eqref{gene-structerm} and \eqref{chermitik} for the following conditions
\begin{eqnarray}
h_\rho(m\bullet f,n\bullet k)&=&\rho(f)^\dag\star h_\rho(m,n)\star\rho(k)\label{twis-hermitor},\\
h_\rho(m,n)^\dag&=&h_\rho(n,m)\label{chermitik1},
\end{eqnarray}
with $h_\rho(1,1)=1$. Note that consistency of \eqref{twis-hermitor} with \eqref{chermitik1} requires the left and right twists to be equal. Assume now that $\rho$ is a regular automorphism, which is the most relevant case in the present situation. Then one infers that 
\begin{equation}
h(f,k)=\rho^{-1}\sigma(f^\dag)\star\rho\sigma^{-1}(k).\label{retwist}
\end{equation}
From this, one easily realizes that the analysis of Subsection \ref{section33} can be reproduced by simply replacing $\sigma$ by $\xi=\rho^{-1}\sigma$. In particular, this would lead to the same special value for the dimension of the $\kappa$-Minkowski space.\\

Some phenomenological consequences related to the action functional \eqref{zelagrangien} have been examined and discussed in \cite{MW20-bis}. In particular, assuming a simple compactification scheme of the predicted extra dimension on the orbifold $\mathbb{S}^1/\mathbb{Z}_2$ in the spirit of the models with Universal Extra Dimension (UED), one finds that consistency with the recent LHC data on the size $\mu^{-1}$ of the extra dimension ($\mu\gtrsim\mathcal{O}(1-5)\ \text{TeV}$) requires 
\begin{equation}
\kappa\gtrsim\mathcal{O}(10^{13})\ \text{GeV}\label{kappaconserv}
\end{equation}
upon identifying $\kappa$ with the 5-dimensional bulk Planck mass. In the action functional \eqref{zelagrangien}, the expression for the kinetic operator 
is rigidly fixed by the expression of the curvature together with the twisted differential calculus linked to $ \mathfrak{D}_\gamma$ and implies in particular the appearance of a deformed dispersion relation for the photons, which, expanded in powers of $1/\kappa$, takes the form
\begin{equation}
E^2-|\vec{p}| ^2-\frac{1}{\kappa}E^3+\mathcal{O}(\frac{1}{\kappa^2})=0\label{dispersion}.
\end{equation}
Recent observational constraints stemming from Gamma Ray Bursts data improve the above lower bound \eqref{kappaconserv} as
\begin{equation}
\kappa\gtrsim\mathcal{O}(10^{17}-10^{18})\ \text{GeV}\label{kappa-mgm},
\end{equation}
which, assuming that the usual UED model relation $M_P^2=\kappa^3/\mu$ holds true, as for the determination of \eqref{kappaconserv}, would correspond to a very small extra dimension size, namely $\mu\gtrsim\mathcal{O}(10^{13}-10^{16})\ \text{GeV}$.\\

\section{Conclusions.}\label{section5}

We have considered $\kappa$-Poincar\'e invariant gauge theories on $\kappa$-Minkowski space-time, which are noncommutative analogs of the usual $U(1)$ gauge theory. They are obtained from noncommutative twisted connections defined on a hermitian right module, assumed to be a copy of the algebra $\mathcal{M}^d_\kappa$. We have shown that twisting the action of this algebra on the hermitian module does not affect the specific value $d=5$ of the dimension of $\mathcal{M}^d_\kappa$ for which $\kappa$-Poincar\'e invariant gauge theories can exist. The gauge group is as well essentially not changed compared to the case of an untwisted action. Only the usual hermitian condition for the gauge potential $A_\mu=A_\mu^\dag$ can become twisted. As it can be expected, the present framework bears some similarity with some algebraic properties of the twisted spectral triples, used for instance recently in the construction of alternative versions \cite{martinet, Devastato-Martinetti, Filaci-Martinetti} of the Standard Model {\it{``\`a la Connes''}}.\\

Hence, the present analysis enforces the prediction of an extra (spatial) dimension as a characteristic features of the $\kappa$-Poincar\'e invariant gauge theories considered here. One theoretical ground, it would be interesting to examine if this value is changed in $\kappa$-Poincar\'e invariant gauge theories
built from noncommutative connections on a bimodule. One physical ground, it appears that the noncommutative gauge symmetry of the 5-d theory is radiatively broken as shown in \cite{HMW1}. Hence, the mass for the 5-d gauge potential $A_\mu$ may no longer be protected and may receive small but non zero quantum corrections which may also translate to the 4-d photon. The related physical consequences confronted with observational data may well provide constraints on the present framework. We will come back to these points in forthcoming works.

\vskip 2 true cm
{\bf{Acknowledgments}}: Ph. M. thanks the organizers of the conference Geometric Foundations of Gravity, that took place from June 28 to July 2, 2021 in Tartu, Estonia, for giving him the opportunity to give a talk. We all thank the Action CA18108 QG-MM, ``Quantum Gravity Phenomenology in the multi-messengers approach'', from the European Cooperation in Science and Technology (COST). Ph. M. is supported by the NSF grant 1947155 and the JTF grant 61521. J.-C. W. thanks G. Landi and P. Martinetti for past discussions.\\

\appendix
\section{$\kappa$-Poincar\'e algebra and deformed translations.}\label{apendixA}

We use the bicrossproduct basis \cite{majid-ruegg}. $\Delta:\mathcal{P}^d_\kappa\otimes\mathcal{P}^d_\kappa\to\mathcal{P}^d_\kappa$, $\epsilon:\mathcal{P}^d_\kappa\to\mathbb{C}$ and ${\bf{S}}:\mathcal{P}^d_\kappa\to\mathcal{P}^d_\kappa$ are respectively the coproduct, counit and antipode equipping $\mathcal{P}^d_\kappa$ with a Hopf algebra structure. $(P_i, N_i,M_i, \mathcal{E},\mathcal{E}^{-1})$, $i=1, 2, \hdots, d-1$, denote respectively the momenta, boosts, rotations and $\mathcal{E}:=e^{-P_0/\kappa}$. They generate the Lie algebra
\begin{equation}
[M_i,M_j]= i\epsilon_{ij}^{\hspace{5pt}k}M_k,\ [M_i,N_j]=i\epsilon_{ij}^{\hspace{5pt}k}N_k,\ [N_i,N_j]=-i\epsilon_{ij}^{\hspace{5pt}k}M_k\label{poinc1}, 
\end{equation}
\begin{equation}
[M_i,P_j]= i\epsilon_{ij}^{\hspace{5pt}k}P_k,\ [P_i,\mathcal{E}]=[M_i,\mathcal{E}]=0,\ [N_i,\mathcal{E}]=\frac{i}{\kappa}P_i\mathcal{E}\label{poinc2},
\end{equation}
\begin{equation}
[N_i,P_j]=-\frac{i}{2}\delta_{ij}\left(\kappa(1-\mathcal{E}^{2})+\frac{1}{\kappa}\vec{P}^2\right)+\frac{i}{\kappa}P_iP_j\label{poinc3}.
\end{equation}
The Hopf algebra structure is defined by
\begin{align}
\Delta P_0&=P_0\otimes\mathbb{I}+\mathbb{I}\otimes P_0,\ \Delta P_i=P_i\otimes\mathbb{I}+\mathcal{E}\otimes P_i,\ \Delta \mathcal{E}=\mathcal{E}\otimes\mathcal{E}\label{hopf1},\\
\Delta M_i&=M_i\otimes\mathbb{I}+\mathbb{I}\otimes M_i,\ \Delta N_i=N_i\otimes \mathbb{I}+\mathcal{E}\otimes N_i-\frac{1}{\kappa}\epsilon_{i}^{\hspace{2pt}jk}P_j\otimes M_k,\label{hopf2}\\
\epsilon(P_0)&=\epsilon(P_i)=\epsilon(M_i)=\epsilon(N_i)=0,\  \epsilon(\mathcal{E})=1\label{hopf3},\\
{\bf{S}}(P_0)&=-P_0,\ {\bf{S}}(\mathcal{E})=\mathcal{E}^{-1},\  {\bf{S}}(P_i)=-\mathcal{E}^{-1}P_i,\  {\bf{S}}(M_i)=-M_i,\\
{\bf{S}}(N_i)&=-\mathcal{E}^{-1}(N_i-\frac{1}{\kappa}\epsilon_{i}^{\hspace{2pt}jk}P_jM_k)\label{hopf4bis}.
\end{align}
The $\kappa$-Minkowski space $\mathcal{M}_\kappa^d$ can be viewed as the dual of the Hopf subalgebra $\mathcal{T}_\kappa^d$ generated by $P_\mu$, $\mathcal{E}$, the so-called deformed translation algebra. It has a structure of $^*$-Hopf algebra through: $P_\mu^\dag=P_\mu$, $\mathcal{E}^\dag=\mathcal{E}$. Then, the following relation holds true
\begin{equation}
(t\triangleright f)^\dag={\bf{S}}(t)^\dag\triangleright f^\dag,\label{pairing-involution}
\end{equation}
for any $t$ in $\mathcal{T}_\kappa^d$, $f\in\mathcal{M}^d_\kappa$. This yields
\begin{equation}
(P_0\triangleright f)^\dag=-P_0\triangleright(f^\dag),\ (P_i\triangleright f)^\dag=-\mathcal{E}^{-1}P_i\triangleright(f^\dag),\ (\mathcal{E}\triangleright f)^\dag=\mathcal{E}^{-1}\triangleright(f^\dag)\label{dag-hopfoperat}.
\end{equation}
The action of $\mathcal{T}_\kappa^d$ on $\mathcal{M}_\kappa^d$ is $(\mathcal{E}\triangleright f)(x)=f(x_0+\frac{i}{\kappa},\vec{x}),\ \ 
(P_\mu\triangleright f)(x)=-i(\partial_\mu f)(x)$.


\begin{thebibliography}{00}

\bibitem{dopplic}S. Doplicher, K. Fredenhagen, J.E. Roberts, ``{\it{The quantum structure of spacetime at the Planck scale and quantum fields}}'', Commun. Math. Phys. {\bf{172}}(1995), 187. S. Doplicher, K. Fredenhagen and J. E. Roberts, ``{\it{Space-time quantization induced by classical gravity}}'', {Phys. Lett. B\textbf{331}, 39--44 (1994)}. 

\bibitem{amelin} For a comprehensive review, see G. Amelino-Camelia, ``{\it{Quantum Spacetime Phenomenology}}'', Living Rev.Rel. {\bf{16}} (2013) 5.  See also S. Hossenfelder, ``{\it{Minimal Length Scale Scenarios for Quantum Gravity}}'', Living Rev. Rel. {\bf{16}} (2013) 2.

\bibitem{luk2} For a recent review, see J. Lukierski, ``{\it{kappa-Deformations: Historical Developments and Recent Results}}'', J. Phys. Conf. Ser. \textbf{804}, 012028 (2017).

\bibitem{luk1} J. Lukierski, H. Ruegg, A. Nowicki, V. N. Tolsto\"i, ``{\it{$q$-deformation of Poincar\'e algebra}}'', Phys. Lett. B{\bf{264}} (1991) 331. J. Lukierski, A. Nowicki, H. Ruegg, ``{\it{New quantum Poincar\'e algebra and $\kappa$-deformed field theory}}'', Phys. Lett. B{\bf{293}} (1992) 344.

\bibitem{leningrad} V. G. Drinfeld, ``{\it{Quantum Groups}}'', in Proc. Int. Cong. Math., Vols 1,2 (Berkeley 1986) AMS, Providence, RI (1987) 798. 
L. A. Takhtadzhyan, ``{\it{Lectures on quantum groups}}'', Nankai Lectures on Mathematical Physics, Mo-Lin-Ge and Bao-Heng-Zhao Eds., World Scientific (1989).



\bibitem{ame-ca1} G. Amelino-Camelia, ``{\it{Doubly special relativity}}'', Nature {\bf{418}} (2002) 34. G. Amelino-Camelia, G. Gubitosi, A. Marciano, P. Martinetti, F. Mercati, ``{\it{A no-pure boost uncertainity principle from spacetime noncommutativity}}'', Phys. Lett. B{\bf{671}} (2009) 298. For a review on Doubly Special Relativity, see e.g J. Kowalski-Glikman, ``{\it{Introduction to dsr}}'' in Planck scale Effects in Astrophysics and Cosmology, Lecture Notes in Phys. {\bf{669}} (Springer, Berlin 2005) 131, and references therein.

\bibitem{AC1} G. Amelino-Camelia, ``{\it{Testable scenario for Relativity with minimum-length}}'', Phys. Lett. B{\bf{510}} (2001) 255. J. Kowalski-Glikman, ``{\it{Introduction to Doubly Special Relativity}}'', Lect. Notes Phys. {\bf{669}} (2005) 131.

\bibitem{majid-ruegg} S. Majid and H. Ruegg, ``{\it{Bicrossproduct structure of $\kappa$-Poincar\'e group and non-commutative geometry}}'', Phys. Lett. B{\bf{334}} (1994) 348.

\bibitem{habsb-impbis} M. Dimitrijevi\'c, L. Jonke, L. M\"oller, E. Tsouchnika, J. Wess, M. Wohlgennant ``{\it{Deformed field theory on $\kappa$-spacetime}}'', Eur. Phys. J. C{\bf{31}} (2003) 129. 

\bibitem{ital-1bis} A. Agostini, G. Amelino-Camelia, M. Arzano, F. D'Andrea, ``{\it{Action functional for kappa-Minkowski noncommutative spacetime}}'', [arxiv:hep-th/0407227]. See also A. Agostini, G. Amelino-Camelia, F. D'Andrea, ``{\it{Hopf-algebra description of noncommutative-spacetime symmetries}}'', Int.J.Mod.Phys. A{\bf{19}} (2004) 5187. 

\bibitem{ital-2bis} A. Agostini, G. Amelino-Camelia, M. Arzano, A. Marciano, R. Altair Tacchi, ``{\it{Generalizing the Noether theorem for Hopf-algebra spacetime symmetries}}'', Mod.Phys.Lett.A{\bf{22}} (2007) 1779. G. Amelino-Camelia, M. Arzano, {\it{``Coproduct and star-product in field theories on Lie-algebra noncommutative space-times''}}, {Phys. Rev. D {\bf{65}}} (2002) 084044.

\bibitem{hrvat-1} S. Meljanac, A. Samsarov, ``{\it{Scalar field theory on kappa-Minkowski spacetime and translation and Lorentz invariance}}'', Int. J. Mod. Phys. A{\bf{26}} (2011) 1439. E. Harikunmar, T. Juri\'c, S. Meljanac, ``{\it{Electrodynamics on $\kappa$-Minkowski space-time}}'', Phys. Rev. D{\bf{84}} (2011) 085020. S. Meljanac, A. Samsarov, J. Trampetic, M. Wohlgenannt, ``{\it{Scalar field propagation in the $\phi^4$ kappa-Minkowski model}}'', JHEP {\bf{12}} (2011) 010. H. Grosse, M. Wohlgenannt, ``{\it{On $\kappa$-Deformation and UV/IR Mixing}}'', Nucl.Phys. B{\bf{748}} (2006) 473.

\bibitem{mercati1} F. Mercati and M. Sergola, ``{\it{ Pauli-Jordan Function and Scalar Field Quantization in $\kappa$-Minkowski Noncommutative Spacetime}}'', Phys. Rev. D{\bf{98}} (2018) 045017.

\bibitem{PW2018} T. Poulain, J.-C.Wallet, {\it{``$\kappa$-Poincar\'e invariant quantum field theories with KMS weight''}}, Phys. Rev. D{\bf{98}} (2018) 025002.

\bibitem{PW2018bis} T. Poulain, J.-C. Wallet, {\it{``$\kappa$-Poincar\'e invariant orientable field theories at 1-loop''}}, JHEP {\bf{01}} (2019) 064.

\bibitem{Grosse-1}
H.~Grosse and R.~Wulkenhaar, ``{\it{Renormalisation of $\varphi^4$-theory on noncommutative $\mathbb{R}^2$ in the matrix base}}'', JHEP {\bf 0312} (2003) 019. H.~Grosse and R.~Wulkenhaar, ``{\it{Renormalisation of $\varphi^4$-theory on noncommutative $\mathbb{R}^4$ in the matrix base}}'', Commun.\ Math.\ Phys.\  {\bf 256} (2005) 305. H. Grosse, R. Wulkenhaar, ``{\it{Self-dual noncommutative $\varphi^4$-theory in four dimensions is a non-perturbatively solvable and non-trivial quantum field theory}}'', Commun. Math. Phys. {\bf{329}} (2014) 1069. 

\bibitem{wal-moyal1} H.~Grosse and M.~Wohlgenannt, ``\textit{Induced gauge theory on a noncommutative
  space}'', Eur. Phys. J. \textbf{C52} (2007) 435.
A.~de~Goursac, J.-C. Wallet and R.~Wulkenhaar, ``\textit{Noncommutative induced
  gauge theory}'', Eur. Phys. J. \textbf{C51} (2007) 977.
J.-C. Wallet, ``{\it{Noncommutative Induced Gauge Theories on Moyal Spaces}}'', {J. Phys. Conf. Ser. \textbf{103}, 012007 (2008)}. %
See also P. Martinetti, P. Vitale, J.-C. Wallet, ``{\it{Noncommutative gauge theories on $\mathbb{R}^2_\theta$ as matrix models}}'', {JHEP} \textbf{09} (2013) 051.
\bibitem{jncg1}
E. Cagnache, T. Masson and J-C. Wallet, ``{\it{Noncommutative Yang-Mills-Higgs actions from derivation based differential calculus}}'', {J. Noncommut. Geom. \textbf{5}, 39--67 (2011)}. D. N. Blaschke, H. Grosse, J.-C. Wallet, ``{\it{Slavnov-Taylor identities, non-commutative gauge theories and infrared divergences}}'', JHEP {\bf{06}} (2013) 038.
\bibitem{wal-moyal11}
A. de Goursac, J.-C. Wallet, R. Wulkenhaar, ``{\it{On the vacuum states for noncommutative gauge theory}}'', {Eur. Phys. J. C\textbf{56} (2008) 293--304}. %
See also A. de Goursac, A. Tanasa, J.-C. Wallet, ``{\it{Vacuum configurations for renormalizable noncommutative scalar models}}'', Eur. Phys. J. C{\bf{53}} (2008) 459.


\bibitem{vign-sym} F. Vignes-Tourneret, ``{\it{Renormalisation of the orientable noncommutative Gross-Neveu model}}'', Ann. H. Poincar\'e {\bf{8}} (2007) 427. 
A. de Goursac, J.-C. Wallet, ``{\it{Symmetries of noncommutative scalar field theory}}'', J.\ Phys.\ A: Math. Theor. {\bf{44}} (2011) 055401. 
J.-C. Wallet, ``{\it{Connes distance by examples: Homothetic spectral metric spaces}}'', Rev. Math. Phys. {\bf{24}} (2012) 1250027.

\bibitem{wal-16} J.-C. Wallet, ``{\it{Exact Partition Functions for Gauge Theories on $\mathbb{R}^3_\lambda$}}'', Nucl. Phys. B{\bf{912}} (2016) 354.
A. G\'er\'e, T. Juri\'c and J.-C. Wallet, ``{\it{Noncommutative gauge theories on $\mathbb{R}^3_\lambda$: Perturbatively 
finite models}}'', {{JHEP} \textbf{12} (2015) 045}.
 P. Vitale, J.-C. Wallet, ``{\it{Noncommutative field theories on $\mathbb{R}^3_\lambda$: Toward UV/IR mixing freedom}}'', {JHEP} \textbf{04} (2013) 115.

\bibitem{DS}  B. Durhuus, A. Sitarz,  ``{\it{Star product realizations of kappa-Minkowski space}}'', J. Noncommut. Geom. \textbf{7} (2013) 605.

\bibitem{old1} J. von Neumann, ``{\it{Die Eindeutigkeit der Schr\"odingerschen Operatoren }}'', Math. Ann. 104 (1931) 570.

\bibitem{old2} H. Weyl, ``{\it{Quantenmechanik und Gruppentheorie}}'', Zeitschrift f\"ur Physik 46 (1927) 1.

\bibitem{poulinerie} T. Juri\'c, T. Poulain, J.-C. Wallet, ``{\it{Involutive representations of coordinate algebras and quantum spaces}}'',  JHEP {\bf{07}} (2017) 116.  T. Poulain, J.-C. Wallet, ``{\it{Quantum spaces, central extensions of Lie groups and related quantum field theories}}'',  J. Phys.: Conf. Ser. 965 (2018) 012032. See also first of ref. \cite{wal-16}.

\bibitem{dimitriev} For early works about gauge theories on $\kappa$-Minkowski, see M. Dimitrijevi\'c, L. Jonke, L. M\"oller, ``{\it{$U(1)$ gauge field theory on kappa-Minkowski space }}'', JHEP {\bf{09}} (2005) 068. M. Dimitrijevi\'c, F. Meyer, L. M\"oller, J. Wess, ``{\it{Gauge theories on the kappa-Minkowski spacetime }}'', Eur.Phys.J.C{\bf{36}} (2004) 117.

\bibitem{rev-gauge}  For a review focused on the use of Drinfeld's twists and related concepts in gauge theories, see 
M. Dimitrijevi\'c, L. Jonke, A. Pachol, ``{\it{Gauge Theory on Twisted $\kappa$-Minkowski: Old Problems
and Possible Solutions}}'', SIGMA {\bf{10}} (2014) 063

\bibitem{MW20} P. Mathieu, J.-C. Wallet, ``{\it{Gauge theories on $\kappa$-Minkowski spaces: Twist and modular operators }}'', JHEP {\bf{05}}(2020) 115.

\bibitem{MW20-bis} Ph. Mathieu, J.-C. Wallet, ``{\it{Single Extra Dimension from $\kappa$-Poincar\'e and Gauge Invariance }}'', JHEP {\bf{03}} (2021) 209.

\bibitem{brst} Ph. Mathieu, J.-C. Wallet, ``{\it{Twisted BRST symmetry in gauge theories on $\kappa$-Minkowski  }}'', Phys. Rev. D {\bf{103}} (2021) 086018.

\bibitem{HMW1} K. Hersent, Ph. Mathieu, J.-C. Wallet, ``{\it{Quantum stability of gauge theories on $\kappa$-Minkowski space}}'',  arXiv:2107.14462 (2021).

\bibitem{como-1} A. Connes, H. Moscovici, ``{\it{Type III and spectral triples}}'', in Traces in number theory,geometry and quantum fields, Aspects of Math. E38, Vieweg, Wiesbaden 2008, pp 57.

\bibitem{martinet} G. Landi, P. Martinetti, ``{\it{Gauge transformations for twisted spectral triples }}'', Lett. Math. Phys. {\bf{108}} (2018) 2589. See also P. Martinetti, J. Zanchettin, ``{\it{Twisted Spectral Triples without the First-Order Condition }}'', arxiv: 2103.15643 (2021).

\bibitem{alaconnes} See A. H. Chamseddine, A. Connes, W.D. van Suijlekom, ``{\it{Beyond the spectral standard model:
emergence of Pati-Salam unification}}'', JHEP {\bf{11}} (2013) 132 and references therein.

\bibitem{takesaki} For a comprehensive exposition of the Tomita-Takesaki theory, see M. Takesaki, ``{\it{Theory of Operator Algebras I-III}}'', EMS Vols. 124, 125, 127, Springer 2002.

\bibitem{Matassa2012} M. Matassa, ``{\it{A modular spectral triple for $\kappa$-Minkowski space}}'',  J. Geom. Phys. {\bf{76}} (20141) 025011.

\bibitem{Matassa2013} M. Matassa, ``{\it{On the spectral and homological dimension of $\kappa$-Minkowski space}}'', arxiv: 1309.1054 (2013).

\bibitem{Devastato-Martinetti} A. Devastato, P. Martinetti, ``{\it{Twisted Spectral Triple for the Standard Model and Spontaneous Breaking of the Grand Symmetry}}'', Math. Phys. Anal. Geom. {\bf{20}}, 2 (2017).

\bibitem{Filaci-Martinetti} M. Filaci, P. Martinetti, S. Pesco, ``{\it{Minimal twist for the Standard Model in noncommutative geometry : the field content}}'',  Phys. Rev. D {\bf{104}} (2021) 025011.

\end{thebibliography}
\end{document}